\documentclass[10pt, conference, letterpaper]{IEEEtran}
\IEEEoverridecommandlockouts
\usepackage{cite}
\usepackage{amsmath,amssymb,amsfonts}
\usepackage{algorithmic}
\usepackage{graphicx}
\usepackage{textcomp}
\usepackage{xcolor}
\usepackage{multirow}

\usepackage[normalem]{ulem}
\usepackage{xspace}
\usepackage{tabularx}
\usepackage{array}
\usepackage{caption}
\usepackage[utf8]{inputenc}

\newcommand\name{BodyPIN\xspace}

\newcommand\ie{\textit{i.e.,}\xspace}
\newcommand\etc{\textit{etc}\xspace}
\newcommand\eg{\textit{e.g.,}\xspace}

\newcommand{\Figure}[1]{Fig.~\ref{fig:#1}}
\newcommand{\Equation}[1]{Eq.~\ref{eq:#1}}
\newcommand{\Sec}[1]{\S\ref{sec:#1}}
\newcommand{\Table}[1]{Table~\ref{tab:#1}}

\begin{document}

\newcommand{\systemname}{BodyPIN\xspace}

\title{Continuous User Authentication by Contactless Wireless Sensing}


\author{\IEEEauthorblockN{Fei Wang}
\IEEEauthorblockA{\textit{Xi'an Jiaotong University} \\
fai.er@stu.xjtu.edu.cn}
\and
\IEEEauthorblockN{Zhenjiang Li}
\IEEEauthorblockA{\textit{City University of Hong Kong} \\
zhenjiang.li@cityu.edu.hk}
\and
\IEEEauthorblockN{Jinsong Han}
\IEEEauthorblockA{\textit{Zhejiang University} \\
hanjinsong@zju.edu.cn}
}

\maketitle

\begin{abstract}
This paper presents \systemname, which is a continuous user authentication system by contactless wireless sensing using commodity Wi-Fi. \systemname can track the current user's legal identity throughout a computer system's execution. In case the authentication fails, the consequent accesses will be denied to protect the system. The recent rich wireless-based user identification designs cannot be applied to \systemname directly, because they identify a user's various activities, rather than the user herself. The enforced to be performed activities can thus interrupt the user's operations on the system, highly inconvenient and not user-friendly. In this paper, we leverage the bio-electromagneetics domain human model for quantifying the impact of human body on the bypassing Wi-Fi signals and deriving the component that indicates a user's identity. Then we extract suitable Wi-Fi signal features to fully represent such an identity component, based on which we fulfill the continuous user authentication design. We implement a \systemname prototype by commodity Wi-Fi NICs without any extra or dedicated wireless hardware. We show that \systemname achieves promising authentication performances, which is also lightweight and robust under various practical settings.
\end{abstract}


\section{Introduction}
\label{sec:intro}

\begin{figure}[t]
	\includegraphics[width=0.95\linewidth]{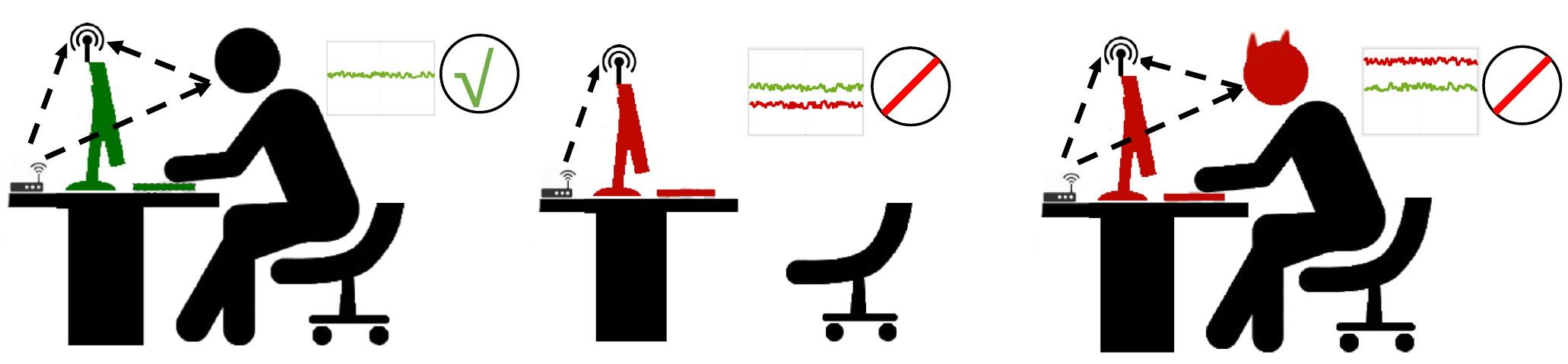}
	\centering
	\caption{\textbf{System illustration}. \textbf{(left)}: Wi-Fi related biometrics are registered by \systemname after a legal user logs in, and the computer can be continuously accessed; \textbf{(middle)}: if the user leaves, the system is unaccessible; \textbf{(right)}: once an adversary appears, biometrics mismatch and adversary's access is denied.
	}
	\label{fig:overview}
	\vspace{-.2in}
\end{figure}

Most computer systems require the user authentication only at the login step. The systems then can be accessed once the authentication is successful, even the user may temporarily leave afterwards~\cite{lin2017cardiac}. 
However, such an \textit{one-time authentication} scheme could expose systems to adversaries, especially during the user's absent period, and cause severe security issues, such as the illegal copy of private documents, the peep of sensitive information, malicious modifications of system configurations, \etc. The victim systems can be common computers, as shown in Fig.~\ref{fig:overview}, and could also be the emerging mobile devices~\cite{feng2017continuous}, \eg smart phones or wearables, as well as various Internet of Things (IoTs) devices in a smart cyber-space.~

To defend this crucial security issue, the concept of \textit{continuous authentication} was proposed~\cite{traore2011continuous}, aiming to keep track of the current user's legal identity throughout the system's operation. In case the authentication interrupts, \eg the legal user leaves and/or the adversary appears, the system is locked automatically. One naive way for achieving this is to ask the user to frequently authenticate herself, \eg by her password or fingerprint, but this will interrupt the user's normal operations on the system, \ie highly inconvenient and not user-friendly.

To overcome this limitation, the \textit{contactless sensing} based designs are widely proposed. Specifically, various sensors can be adopted to sense the user's certain biometric features~\cite{niinuma2010continuous}. Then we can match them with the pre-recorded feature profiles for the authentication. As the entire process is fully \textit{passive} to the user and does not require any user's touch on the device, \eg no password input, the authentication thus 
can continuous, without interrupting the user's operations on the system.

Following this principle, there are two main types of designs proposed in the literature, the camera-based and wireless-based solutions. For the former category, the features, like the colors of the user's cloth and skin~
\cite{niinuma2010continuous} 
and the gaze moving pattern \cite{eberz2015preventing}, can be utilized. However, the camera-based designs suffer two obvious issues. First, due to the limited camera view angle, the user should be directly captured by the camera (without blocking) in good lighting conditions, which may limit the usable scenarios \cite{lin2017cardiac}. Second, cameras can also cause the privacy leakage issue~\cite{cheng2017homespy,hoyle2015sensitive} 
if the recorded video is not properly protected or gets hacked by the adversaries. Due to these concerns, the wireless sensing based designs promising appear recently~\cite{lin2017cardiac,lv2017wii,xin2016freesense,zeng2016wiwho,zhang2016wifi}, which can effectively bypass these two issues. However, the existing designs either require the user to perform certain activities~\cite{xin2016freesense,zeng2016wiwho,zhang2016wifi}, \eg walking, or a dedicated hardware design~\cite{lin2017cardiac}, which could inevitably interrupt the user's operations as well, or increase the system deploying and maintaining costs, \eg not pervasive enough.

Motivated by these existing works, in this paper, we explore the opportunity to achieve the continuous user authentication using commodity wireless techniques, like Wi-Fi~\cite{halperin2011tool}, without imposing any activities performed by the user. If this is viable, the solution should be able to preserve the merits from prior wireless-based designs, and meanwhile also largely reduces the system cost. However, the key questions is \textit{without the dedicated wireless design, whether suitable features from Wi-Fi signals exist to strongly identify a user's identity along for the continuous authentication design}. Such an identification implies that the explored features should be related to the user's biometric features directly, rather than the performed activities as previously studied~\cite{zeng2016wiwho,zhang2016wifi,xin2016freesense}, which, to our best knowledge, has not been explored yet.

Our investigation in this paper is inspired by the existing studies from the bio-electromagnetics domain~\cite{christ2006characterization,dove2014analysis,melia2013electromagnetic,gabriel1996dielectric2}, which have the proper model to abstract the human body for understanding the interactions between the electromagnetic waves and human body. Based on such a design preliminary, we quantify the impact of our body on bypassing Wi-Fi signals and derive the component that indicates a user's identity, which is jointly determined by the user body's appearance, \eg the radius of our body's intersecting surface, and also our body's internal factors, \eg permittivity, permeability, body-fat ratio, \etc. The component is hence highly user-dependent, which is qualified for the user authentication. Then, our next effort is to extract suitable Wi-Fi signal features to fully represent the user's identity component. To this end, we conduct an in-depth analysis and figure out a set of  Wi-Fi biometrics traits or features from the channel state information (CSI)~\cite{shen2006channel}. Based on this, we finally design a continuous authentication system, \systemname, which can achieve both a high true-positive (TP) rate, for the least interruption to the legal users, and a low false-positive (FP) rate, for the least misses of the adversaries.

Fig.~\ref{fig:overview} depicts how \systemname works. In Fig.~\ref{fig:overview}~(left), when a legal user logs in a computer system, her Wi-Fi related biometrics features are registered for the continuous authentication. Later, when the user leaves, the Wi-Fi feature matching becomes unsuccessful and the system turn to be unaccessible, as in Fig.~\ref{fig:overview}(middle). In such a case, the system can deny the access from those who have the mismatched Wi-Fi biometrics features, as illustrated in Fig.~\ref{fig:overview}~(right). Following this working flow, we implement a \systemname prototype using Intel 5300 wireless NICs. Extensive evaluations show that it can achieve very good authentication performance, 
 nearly 90\% authenticating accuracy and defending precision with a group of 30 subjects. 
The computation is light-weighted, around 300$ms$, which is sufficient for the real-time authentication. 

In summary, the contributions of this paper are as follows:
\begin{itemize}
	\item We propose a continuous user authentication system, \systemname, using the commodity Wi-Fi signals through a contactless wireless sensing design.
	
	\item We identify the signal component that are directly related to each individual user and extract a set of suitable Wi-Fi signal features to represent it for the authentication.
	
	\item We implement a \systemname prototype and conduct extensive experiments for the evaluation, which demonstrates promising and robust performance.
\end{itemize}


\section{Impact of Human Body on Wi-Fi Signals}\label{sec:model}

Existing studies have empirically demonstrate that our human body could have impacts on the electromagnetic waves, like absorption, within a certain frequency band, which covers Wi-Fi's frequencies~\cite{christ2006characterization,dove2014analysis,melia2013electromagnetic,gabriel1996dielectric2}. In this section, we strive to further quantify the impact and setup the relation between the Wi-Fi signal features and each individual user's characteristics,  based on which we can achieve the \systemname design.

\begin{figure}[t]
	\includegraphics[width=1\linewidth]{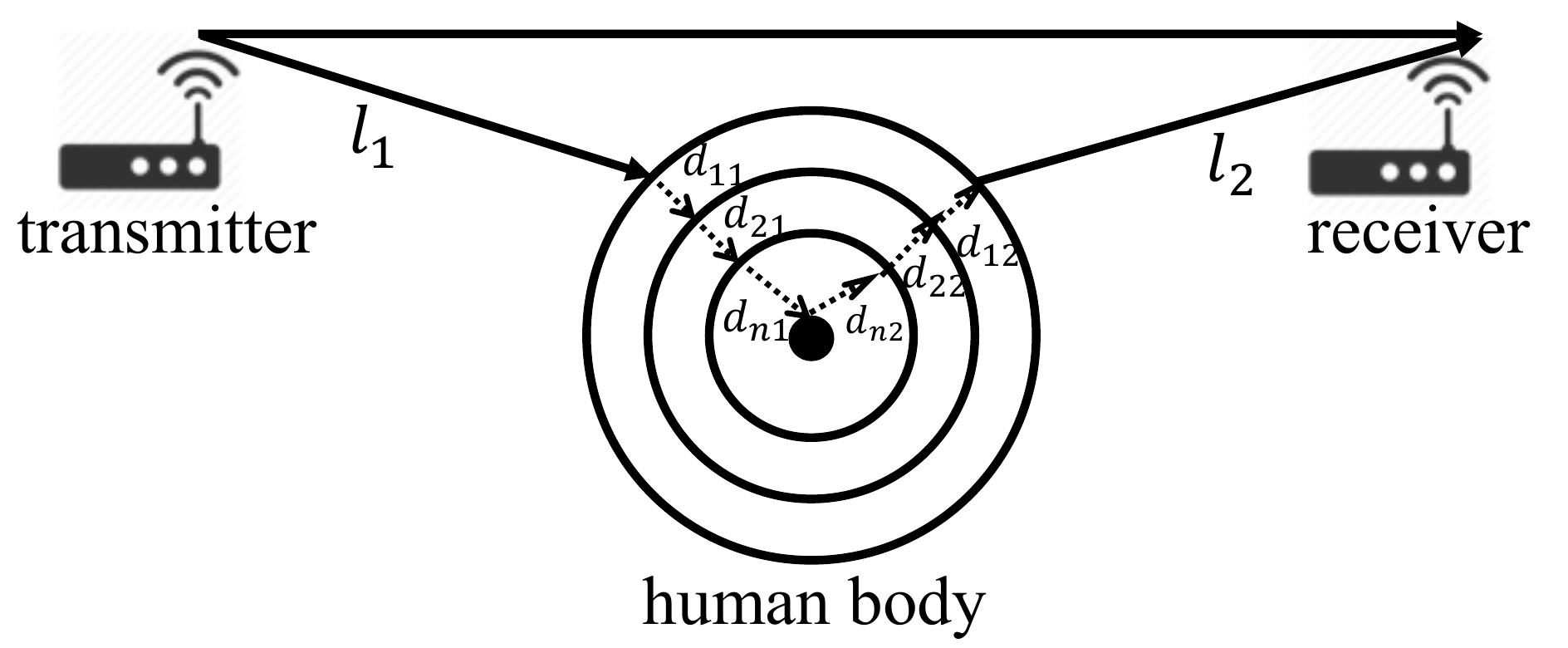}
	\centering
	\caption{The human body is modeled as a series of circles that represent different tissue layers and each layer could cause different attenuations for the bypassing Wi-Fi signals.}
	\label{fig:model}
\end{figure}

To this end, we first borrow the classic human model from the bio-electromagnetics domain, as in \Figure{model}, which abstracts the human body as a series of circles to represent different tissues~\cite{gabriel1996dielectric2}, such as the skin, fat under the skin, muscle, fat on the viscera, viscera, bone, etc., and the radius of layer $i$ is $r_i$, where $i \in[1, n]$ and $n$ is the total number of layers. Although this model is simple, similar as prior studies~\cite{dove2014analysis}, we find that it is effective enough for our analysis (\S\ref{sec:evaluation}).

As shown in \Figure{model}, we denote the distances from the user to the transmitter and receiver as $l_1$ and $l_2$, respectively. Other useful notations are tabulated in \Table{symbols}. According to the table, the initial Wi-Fi signals generated at the transmitter side, \ie $X$, can be mathematically written as:
\begin{eqnarray}
X &=& A \cdot e^{-j \cdot 2\pi \cdot f \cdot t+\phi_0},
\label{eq:s}
\end{eqnarray}
where $\phi_0$ is the initial phase, and $A$ and $f$ are the amplitude and frequency, respectively. After the signal $X$'s transmission from the transmitter to the receiver, its amplitude and phase both will change, and the receiver can receive multiple copies of this signal due to the multipath. For the cope that bypasses the user's body (stated below), we find that its signal changes convey the user's biometric features.

\textbf{Amplitude}. During signal $X$'s transmission, its amplitude $A$ decays, in terms of the signal power, along the time. Such a decay effect occurs for every propagation medium. In particular for the power decay of wireless signals, the decayed power amount can be computed through the Friis transmission equation\cite{shaw2013radiometry}. To facilitate the understanding, we omit the sophisticated intermediate steps and provide the expression of received signal amplitude $A'$ reflected by the user as follows:
\begin{eqnarray}
A' &=& A \cdot c_1^0 \cdot c_2^0 \cdot \prod\nolimits_{i=1}^{n}c_i,
\label{eq:a}
\end{eqnarray}
where $c_i$ is the power decay due to the layer $i$ of the user's body (\Figure{model}), $n$ is the total number of layers that reflect the signals, $c_1^0$ and $c_2^0$ represent power decays from the transmitter to the user and from the user to the receiver, respectively.

\begin{table}[t]
	\centering
	\begin{tabular}{l||l}
		\hline
		\textbf{Symbols} & \textbf{Meaning}                      \\ \hline \hline
		$f$      & Frequency of the Wi-Fi signal    \\ \hline
		$A$      & Initial amplitude of the Wi-Fi signal    \\ \hline
		$\phi_0$      & Initial phase of the Wi-Fi signal    \\ \hline
		$m$ & Total amount of propagation mediums \\\hline
		$l_1$  &  Distance from the transmitter to the user \\ \hline
		$l_2$  &  Distance from the user to the receiver\\ \hline
		$d_{i1}$  &  In-body length of Wi-Fi in the the $i^{\text{th.}}$ layer of human body\\ \hline
		$d_{i2}$  &  Out-body length of Wi-Fi in the the $i^{\text{th.}}$ layer of human body\\ \hline
		$\mu_i$  & Permeability of the $i^{\text{th.}}$ layer of human body \\ \hline
		$ \varepsilon_i$  &  Permittivity of the $i^{\text{th.}}$ layer of human body \\ \hline
		$\mu_0$  & Permeability of the air \\ \hline
		$ \varepsilon_0$  &  Permittivity of the air \\ \hline
		$c_{1}^{0}$      & Power decay from the transmitter to the user \\ \hline
		$c_{2}^{0}$      & Power decay from the user to the receiver\\ \hline
		$c_i$      & Power decay in the $i^{\text{th.}}$ layer of human body \\ \hline 
		
	\end{tabular}
	\caption{List of the mathematical symbols.}
	\label{tab:symbols}
	\vspace{-.15in}
\end{table}

\textbf{Phase}. Next, we compute the phase changes, which are caused by the time (propagation) delay, $t$. In particular, we consider $t$ as a sum of every time delay taking place at every propagation medium, \eg in the air or various body layers, as $t = \sum\nolimits_{i=1}^{m}t_i$, where $m$ is the total amount of propagation mediums and $t_i$ is the delayed time caused by every medium. Each $t_i$ can be calculated by $t_i = \frac{d_i}{v_i}$, where $d_i$ is the length of the $i^{\text{th}}$ medium and $v_i$ is the speed of Wi-Fi signals in this medium. The $v_i$ can be further computed via $v_i = \frac{1}{\sqrt{\mu_i \varepsilon_i}}$, where $\mu_i$ and $\varepsilon_i$ represent the permeability and permittivity of the transmission medium $i$. By combining the three equations above, we can derive the time delay from each Wi-Fi propagation medium by the summarization as follows.
\begin{eqnarray}
t &=& \sum\nolimits_{i=1}^{m}{d_i \cdot \sqrt{\mu_i \varepsilon_i}}. \notag
\end{eqnarray}
According to \Figure{model}, the time delay $t$ can be rephrased as:
\begin{eqnarray}
t &=& \sum_{i=1}^{n}{(d_{i1}+d_{i2})  \cdot \sqrt{\mu_i \varepsilon_i}} + (l_1+l_2) \cdot \sqrt{\mu_0 \varepsilon_0}
\label{eq:t_sum}
\end{eqnarray}
where the former part is caused by our human body and the later part is caused by propagation over the air. Then according to the \Equation{s}, \Equation{a} and \Equation{t_sum}, the reflected Wi-Fi signal copy $Y$ received by the receiver can be represented by:
\begin{eqnarray}
Y = \uwave{\prod_{i=1}^{n}c_i e^{ -j2\pi f\sum\limits_{i=1}^{n}{(d_{i1}+d_{i2})\sqrt{\mu_i \varepsilon_i}}}}  \uline{c_{1}^{0}c_{2}^{0} e^{
		-j2\pi f((l_1+l_2) \sqrt{\mu_0 \varepsilon_0}) }}  X, \notag
\label{eq:all}
\end{eqnarray}
where the component marked with the wave line is uniquely determined by each individual user. 

\textbf{Summary}. Based on the mathematical expressions of this component, we conclude that the properties of human body, such as the absorption ability, permittivity, permeability and the length of each tissue layer, could have joint impacts on the Wi-Fi signals, which are user-dependent. In \cite{gabriel1996dielectric2}, permittivity and permeability of the tissues, such as the muscle, kidney, liver, \textit{etc.}, have been empiriclally measured. People indeed find that different tissues could cause different influences on the bypassing wireless signals. The intuition is clear --- each type of the tissues has unique compositions and could thus lead to a unique influence on the Wi-Fi signals, which motivates our continuous authentication design in the next section.

On the other hand, as the received $Y$ is just one multipath copy from all received signals by the receiver. It implies that when the user is closer to the line of sight path from the sender to the receiver, it is more likely that the user's unique features can be reliably detected. As unveiled in \S\ref{sec:evaluation}, the authentication is robust in multiple real-world scenarios, which is sufficient for the practical usage.


\section{System Design}
\label{sec:design}

\begin{figure*}[t]
	\includegraphics[width=0.9\linewidth]{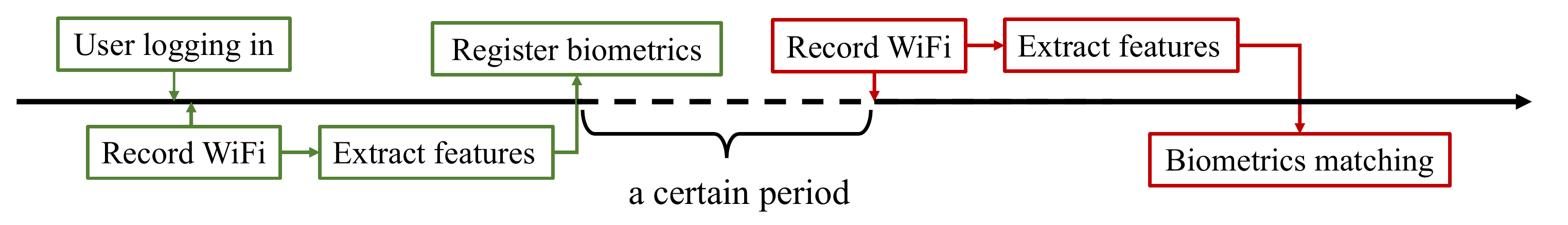}
	\centering
	\caption{System working flow. Once a legal user logs in, the Wi-Fi signals are 
		recorded for the biometric feature extraction. The extracted features are registered for the continuous authentication. To this end, new samples of the Wi-Fi based features are collected and further matched with the registered ones. If matched, the current user is viewed to be legal; Otherwise, \systemname locks the system until the primary authentication is passed again.}
	\label{fig:workflow}
\end{figure*}

In this section, we elaborate the \systemname design. We first describe the system working flow (\S\ref{sec:design:workflow}), followed by the Wi-Fi based biometrics feature extraction (\S\ref{sec:design:feature}) and the authentication design (\S\ref{sec:design:matching}).

\subsection{System working flow}\label{sec:design:workflow}

Fig.~\ref{fig:workflow} shows the working flow of the \systemname system. After a legal user logs into the computer system by any conventional authentication (\ie \textit{primary authentication}), such as passwords, fingerprints, face recognitions, \etc., successfully, \systemname starts to record the Wi-Fi time series. In particular, \systemname processes the channel state information (CSI) from the received Wi-Fi packets, by removing identified amplitude and phase errors, to obtain desired biometric related features. These features (after one- or two-minute recording) are registered in the system and utilized to train a classifier to recognize this legal user for the continuous authentication. More precisely, the system periodically collects CSI samples to generate new Wi-Fi based features about the current user, and then matches them with the registered ones. If matched, the current user is viewed to be legal and the classifier can also be updated by the newly collected features; Otherwise, \systemname locks the system until the primary authentication is passed again.

Two points are worth noting: 1) \systemname is not positioned to replace any primary authentication methods. Hence, the user still needs to well protect their primary authentication keys, like passwords and fingerprints, at the first place. 2) The aim of the on-site feature extraction for training the classifier is to improve the authentication robustness and minimize the possibilities of the false alarm cases that could interrupt the user's normal operations.

\begin{figure*}[t]
	\centering
	\includegraphics[width=.24\linewidth]{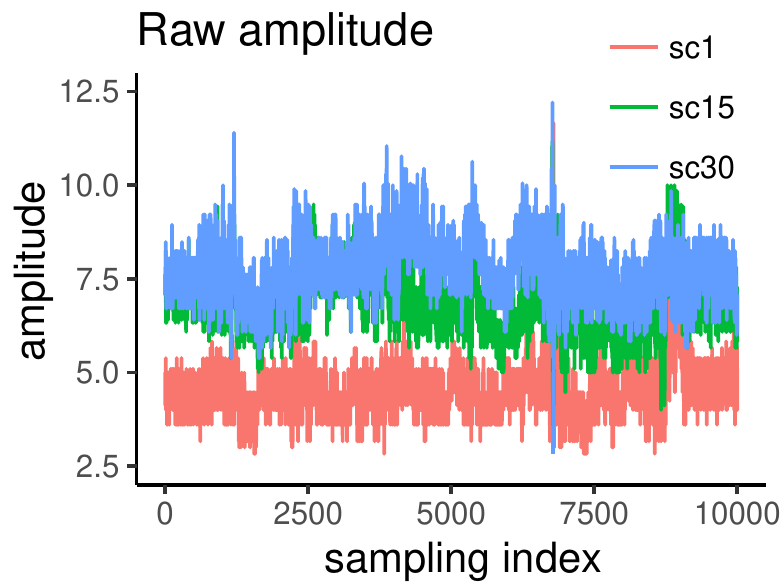}
	\includegraphics[width=.24\linewidth]{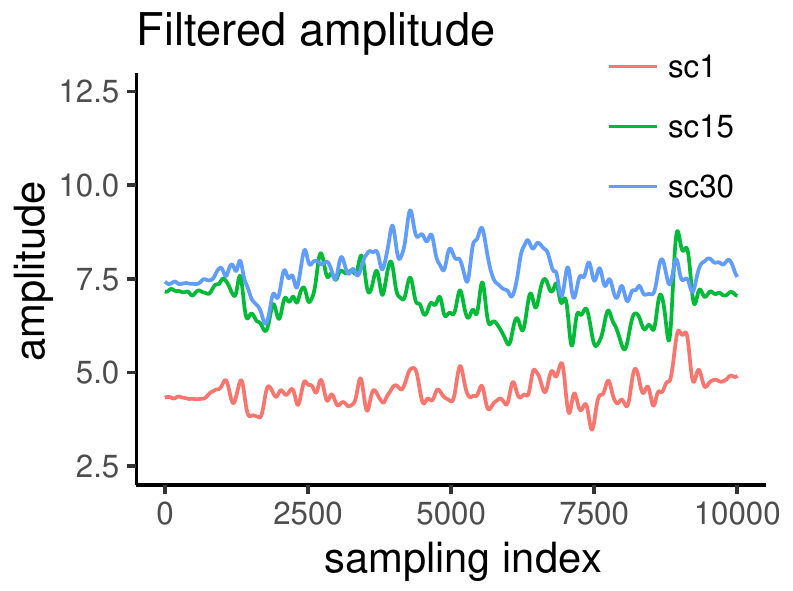}
	\includegraphics[width=.24\linewidth]{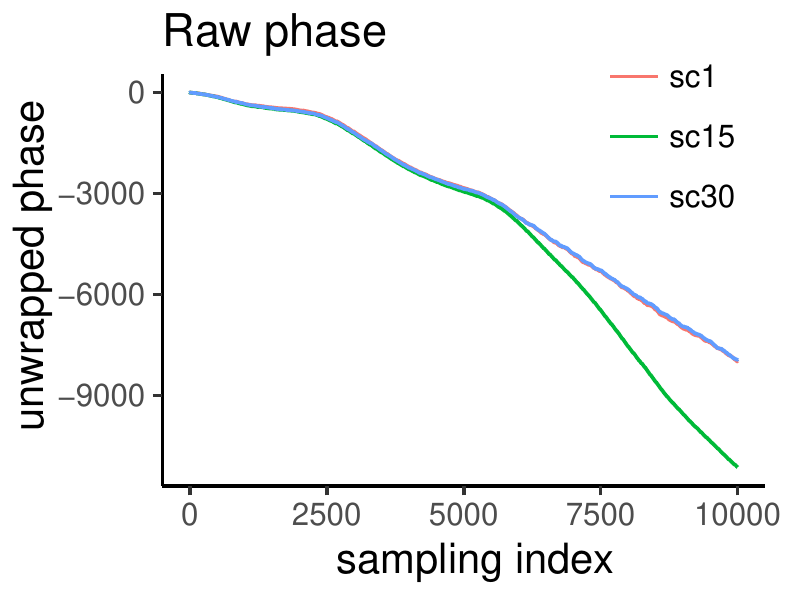}
	\includegraphics[width=.24\linewidth]{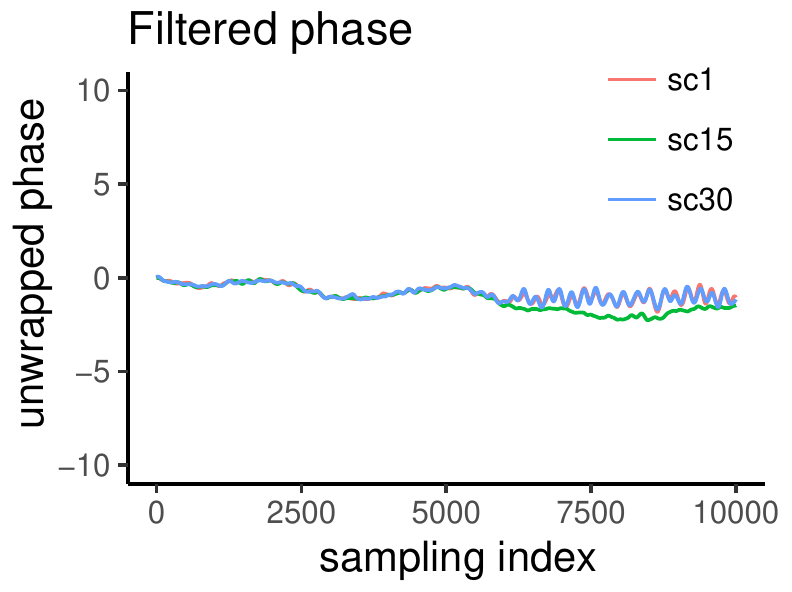}
	\caption{Processing CSI amplitudes and angles. (a), (b): We use low-pass Butterworth filter to suppress the jitters in the CSI amplitude series. (c), (d): We compute and further filter the differences of every two continuous phases. The results are shown in (d), which are for a further feature extraction (series of the $1^{st}$, $15^{th}$ and $30^{th}$ subcarrier are depicted).}
	\label{fig:filter}
\end{figure*}

\subsection{Wi-Fi based biometric feature extraction}
\label{sec:design:feature}

As suggested by the insights from the analysis in Section~\ref{sec:model}, we extract Wi-Fi based user's biometric features in this section. Prior to the design details, we briefly introduce the related Wi-Fi information related to \systemname in the following.

 \textbf{Channel state information}. In modern Wi-Fi protocols, \eg 802.11n/ac, the digit information delivered from the transmitter to the receiver is carried by multiple electromagnetic waves at different frequencies, where each specific-frequency band is called subcarrier, so that the orthogonal frequency division modulation (OFDM)~\cite{nee2000ofdm} can be applied for the data transmission. Supposing the transmitter transmits $X$ and the receiver receives it as $Y$ after the propagation through the wireless channel $H$. We thus have:
\begin{eqnarray}
	Y &=& H \cdot X+n,
\end{eqnarray}
where $n$ is for channel noise. Recently, many advanced Wi-Fi NICs can report the detailed channel state information (CSI) to describe channel $H$ in each subcarrier level, which can be obtained through many existing CSI extraction tools~\cite{halperin2011tool,xie2015precise}. 

 \textbf{Wi-Fi based features}. The CSI information essentially describes the relation of $Y/X$. Further recalling the component marked with the wave line in the derived $Y$ in Section~\ref{sec:model}, we can observe that such a component (related to a series of properties of the user) is also included in the obtained the CSI information, which can be reflected from both the \textit{amplitude} and \textit{phase} two aspects of the CSI.

With many prior investigations~\cite{wang2017wifall,wang2017rt} to extract various types of features from CSI time series, 
in \systemname, we select a preliminary set of features from the CSI amplitude and phase, including  \textit{1) mean, 2) maximum, 3) minimum, 4) mean absolute deviation, (5) interquartile range, (6) root mean square, (7) skewness and (8) kurtosis}. Both amplitudes and phases of all subcarriers, \eg 30 subcarriers from Intel 5300 NICs, can be applied to these features, which lead to the feature dimensions being $8\times30\times2=480$. 

Although rich features could be identified, we find that they cannot be directly adopted, due to the surrounding noises and imperfection of WiFi adapter. As a result, the raw CSI, both amplitude and phase, collected by the CSI tool\cite{halperin2011tool}, will suffer non-negligible fluctuation as illustrated in \Figure{filter}~(a, c). Inspired by the related works~\cite{qian2017inferring,wang2017wifall}, we need to carefully process the collected CSIs (for removing such noises) before designing the classifier for the continuous authentication.

\textit{1) Processing CSI amplitudes}. Generally, when a user sitting before the monitor, her body movement is usually in low frequency. Owing to this, we consider the high-frequency jitters shown in \Figure{filter}~(a) are noises, thus, we apply a low-pass Butterworth filter ($5^{th}$ order, 1Hz of the cut-off frequency) to filter these noises and smooth the time series of CSI amplitude~\cite{qian2017inferring,wang2017wifall}. 
The filtering results are shown in the \Figure{filter}~(b), where the $1^{st}$ and the $2^{nd}$ subfigures are the raw amplitudes and the filtered amplitudes, respectively. We find that the filter can dramatically reduce jitters in the raw amplitude series.

\textit{2) Processing CSI phases}. The noises in the time series of the CSI phases is much different compared with amplitudes, as shown in in \Figure{filter}~(c), it has a decreasing slope in the sampling duration. Prior work have studied this phenomena~\cite{kotaru2015spotfi,xie2015precise,wang2017rt,wang2017biloc} and conclude that it is introduced by joint impacts from a series of offsets shown as follows:
\begin{equation}
\phi = \phi_T + \phi_s + \phi_b + \phi_m + 2\pi f \Delta t,
\label{eq:phase} 
\end{equation}
where $\phi$ and $\phi_T$ stand for the measured phase and true phase, respectively. The $\phi_s$, $\phi_b$ and $\phi_m$ are sampling frequency offset, packet boundary detection uncertainty and measurement error, respectively, which are considered uncontrollable but follow certain probability distributions, \eg the Gaussian. The last component, \ie $2\pi f \Delta t$, is a constant, where $f$ is the carrier frequency offset of the receiver. 

To eliminate the carrier frequency offset in a lightweight manner, we find that the phase errors can be largely removed by the \textit{difference of two continuous phases} as follows:
\begin{equation}
\phi_t' = \phi_{t+1} - \phi_{t},
\label{eq:diffphase}
\end{equation}
where $\phi_t'$ is the phase difference at the sampling time of $t$. We import such differentiated phases, instead of raw phases, to the Butterwork filter for the feature extraction.

\textit{3) Putting them all together}. In summary, after \systemname collects the CSI samples, it first processes their amplitudes and phases, and then extracts the selected features, based on which a classifier can be trained. To avoid the curse of dimensionality,
we apply unsupervised dimensionality reduction on these features by principal component analysis~(PCA)\cite{jolliffe2011principal}. Empirically, we reserve 90\% of information~(variance) in the feature dataset. Prior to train the classifier, we normalize the feature values in the dataset within [-1, +1]. 

\subsection{Continuous authentication via biometrics matching}\label{sec:design:matching}

\begin{figure}[t]
	\centering
	\includegraphics[width=.4\linewidth]{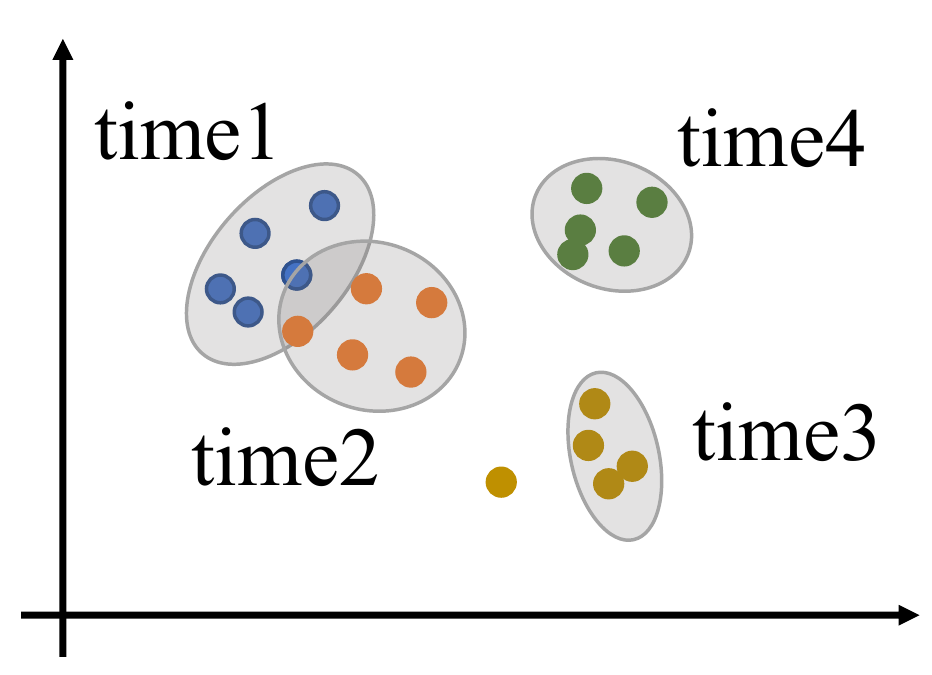}
	\hspace{0.5cm}
	\includegraphics[width=.4\linewidth]{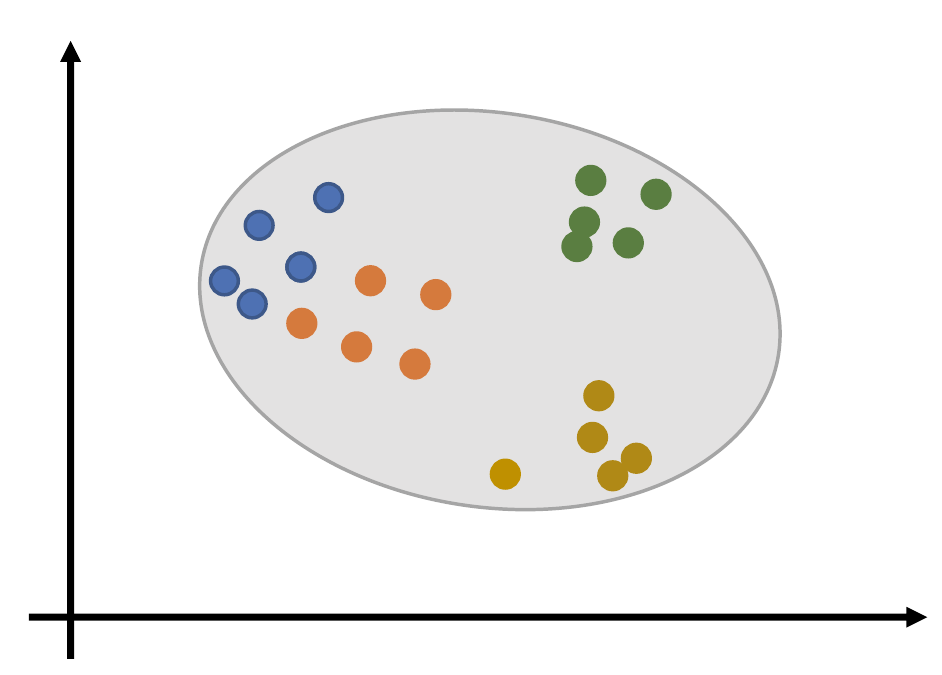}
	\caption{Matching and authentication in a 2D view. \textbf{left}: User's biometrics are extracted at 4 continuous periods and stored. A user whose biometrics features match with one of the stored biometrics is viewed to be legal. \textbf{right}: If we do not use multiple clustering strategy, the authentic range needs to be much larger, which would lead to higher false positive rates.}
	\label{fig:classifier}
\end{figure}

So far, we have introduced the CSI based biometrics feature extraction. In this subsection, we elaborate the matching and authentication designs in \systemname. 

 \textbf{Matching strategy}. According to the system working flow in Fig.~\ref{fig:workflow}, \systemname records the CSI series for registering a legal user's biometrics features when she logs into the computer system for the first time, by the primary authentication. In constructing the classifier to recognize legal users, we consider a practical setting --- as the user may not always stay still, the reflected Wi-Fi signals from the user may vary at the receiver side. As a consequence, the constituents of the user-dependent factors extracted from various signals can be slightly different, \eg the impacts of some factors may vary, even though they belong to the same user.

To tackle this issue, we set \name to continuously record CSI for several periods, \eg each period last for 30 seconds. For example, as shown in the \Figure{classifier}~(left), \name records CSI and extracts biometrics features for 4 periods. In each period, biometrics samples are collected to form a clustering range, and these ranges together can be further converted to an aggregated Bayes probability range, for deciding whether a newly coming feature sample corresponds the legal user. 

Technically, taking the $time1$ shown in \Figure{classifier} for example, in this period, we collect $n$ biometrcis samples, \eg one-second CSI series  contributing one sample, for the legal user, represented by $s_1$ to $s_n$. Supposing the dimension of these samples is $m$, and the value of the $j^{th}$ dimension of the $i^{th}$ sample is represented by $s_{i}^{j}$. With these samples, we first compute the $mean$ , and $variance$ of each dimension, which are represented by $\mu^{j}$ and $\sigma^{2^{j}}$. Afterwards, for any sample $s$, we have its authentic probability equation as follows by the Bayes inference:
\begin{equation}
p(1|s) = \frac{p(1)*p(s|1)}{p(s)},
\label{eq:nbposter}
\end{equation}
where in recording biometrics, $p(1)=1$, and $p(s)$ is unreachable and neglected in our application.
Supposing the values in different dimensions are independently with each other, we have a further equation based on \Equation{nbposter}:
\begin{equation}
p(1|s) \propto  \prod_{j=1}^{m} p(s^j|1)
\label{eq:nbposter2}
\end{equation}

Supposing $p(s^j|1)$ follows a Gaussian distribution. Combining the computed $mean$, $\mu^j$, and $variance$, $\sigma^{2^j}$, we have:
\begin{equation}
p(1|s) \propto  \prod_{j=1}^{m}  \frac{1}{\sqrt{2\pi \sigma^{2^j}}} e^{-\frac{(s^j-\mu^j)^2}{ 2\sigma^{2^j} }}.
\label{eq:nbposter2}
\end{equation}
After the value is computed, we normalize it by an operator of $\sqrt[m]{\cdot}$, and take the result as $p(1|s)$. Finally, we can obtain the probabilities of the $n$ samples, from $p(1|s_1)$ to $p(1|s_n)$. 

 \textbf{Authentication}. For the authentication, we sort these $n$ probabilities in a descending order and set a probability threshold, $p'$, at the 90\%. When facing a new sample, if the probability, $p(1|s_{new})$, is greater than $p'$, \name will consider it is from the legal user. The final decision is jointly made by the probability thresholds at all recording periods: 
\begin{equation}
p(1|s_{new}) \geq p'_{1} \parallel p(1|s_{new}) \geq p'_{2}\parallel  \cdots \parallel   p(1|s_{new})\geq p'_{t}, \notag
\label{eq:decision}
\end{equation}
where $\parallel$ is the operator of the logical OR, and $t$ is the number of continuous CSI based biometrics sampling periods. For the example \Figure{classifier}, the $t$ equals to 4.

\Equation{decision} indicates that if the authentic probability is greater than any one of the probability thresholds, \name considers the user to be legal. As shown in \Figure{classifier}, if the new biometrics sample is within any one of these 4 shadow ranges, it passes the authentication. One significant advantage of this strategy is clear, that is, it makes \name resilient to user state change. In addition, dividing recorded CSI series with multiple smaller periods, comparing with the whole series, can largely decrease the \textit{false positive}~(FP) rate, \ie recognizing an illegal user as the legal one. We still use \Figure{classifier}~(left) to explain this issue. With our strategy, the authentic ranges are 4 small shadow circles. However, if we adopt a long recording time, in order to cover these samples shown in the figure, we need one much larger circle, shown in \Figure{classifier}~(right), which would cover much bigger range and causes higher FP rates.

\textbf{Updating classifier.}~Considering the common variation of user's pose and position, which may lead to new user biometrics features and cause false negative~(FN), we update the classifier with the latest recorded biometrics features during the continuous authentication.

\section{Evaluation}\label{sec:evaluation}

\subsection{Experimental setup}

\begin{figure}[t]
\centering
 \includegraphics[width=0.95\linewidth]{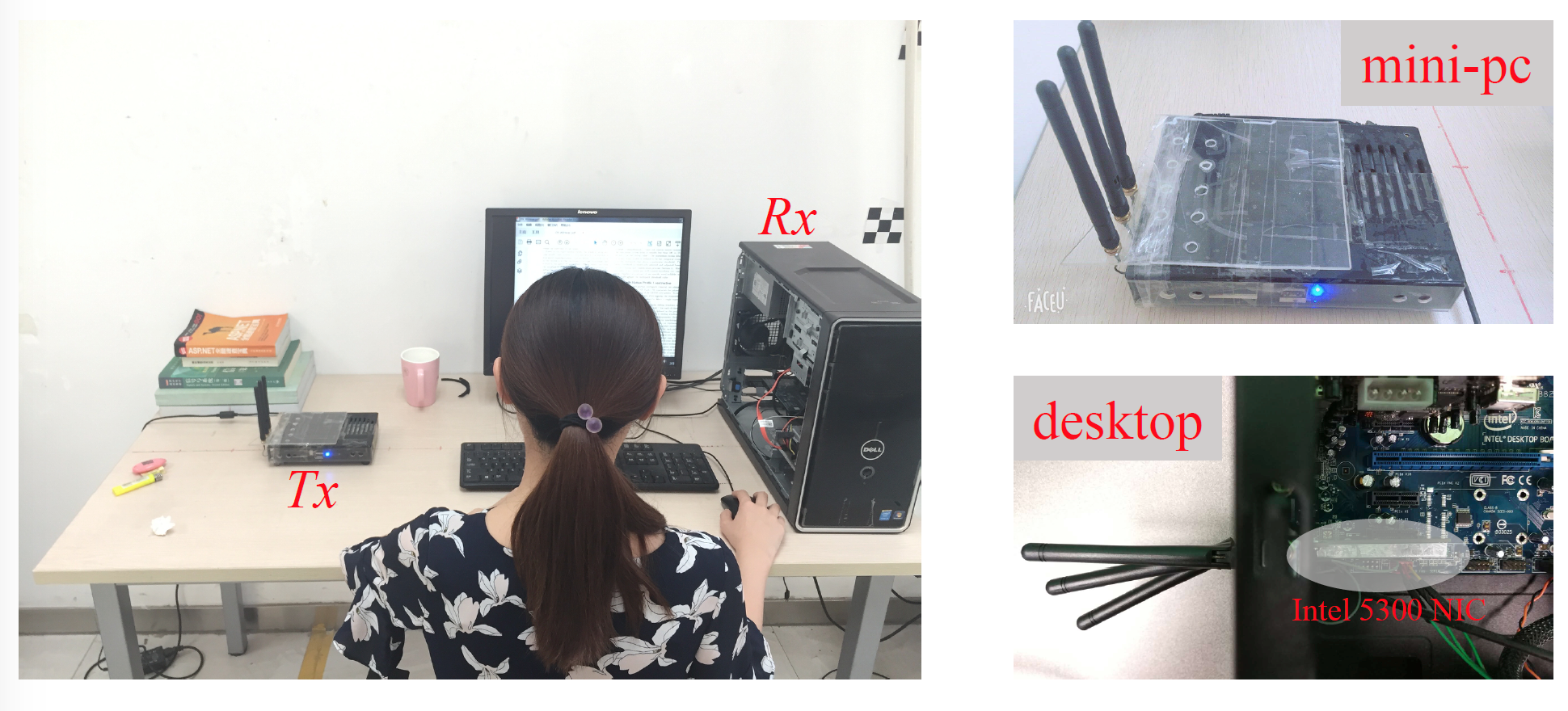} 
\caption{System deployment. We refit a mini-pc with Intel 5300 NIC and take it as the transmitter. A desktop attached with Intel 5300 NIC works as a computer system embedded with \name. Subjects are asked to sit before the monitor and to act as their usual behaviors.}
\label{fig:deployment}
\end{figure}

\begin{figure}[t]
\begin{minipage}{0.5\linewidth}
\centering
\includegraphics[width=1\linewidth]{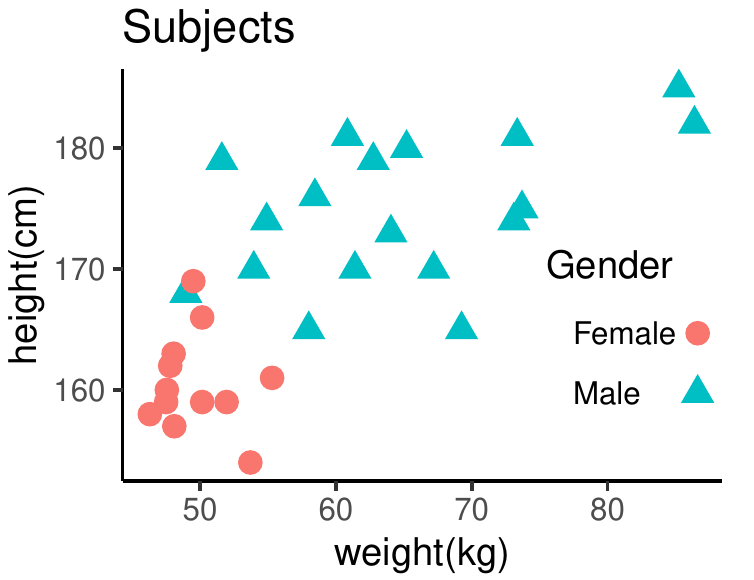}
\end{minipage}
\begin{minipage}{0.4\linewidth}
\centering
\resizebox{3.5cm}{!}{
\begin{tabular}{l||l}
\hline
\textbf{Apparel} & \textbf{Number} \\ \hline\hline
T-shirt    & 14     \\ \hline
Blouse     & 6      \\ \hline
Coat       & 5      \\ \hline
Dress      & 3      \\ \hline
Jacket     & 2      \\ \hline
\end{tabular}
}
\end{minipage}
\caption{Subjects' detailed information. \textbf{(left)}:~Their weights and heights; \textbf{(right)}:~ Subjects' apparels.}
\label{fig:subject}
\end{figure}

In our experiments, we utilize Intel 5300 wireless NICs to record CSI. Specifically, the frequency is 5 GHz and the packet transmission rate is 50 Hz. As shown in \Figure{deployment}, the transmitter is a mini-pc and the receiver is a desktop, which runs Ubuntu 14.04 OS. We use a PCIe-X1 to mini-PCIe adapter to make the card attached on the motherboard of desktop. 30 subjects are recruited in the experiment. Some of them play as the common user and operate on the computer. Other subjects act as the surrounding people to mainly investigate the robustness of our system against such an influence. The detailed information of these 30 subjects, \ie their weights, heights and apparels, is recorded in \Figure{subject}.

\subsection{Overall performance}\label{sec:primary}

We show overall performances, including \textit{mean interruption interval, mean authentication accuracy, mean defending precision and authentication time delay} in this subsection. 
\begin{figure*}[t]
\centering
\includegraphics[width=.3\linewidth]{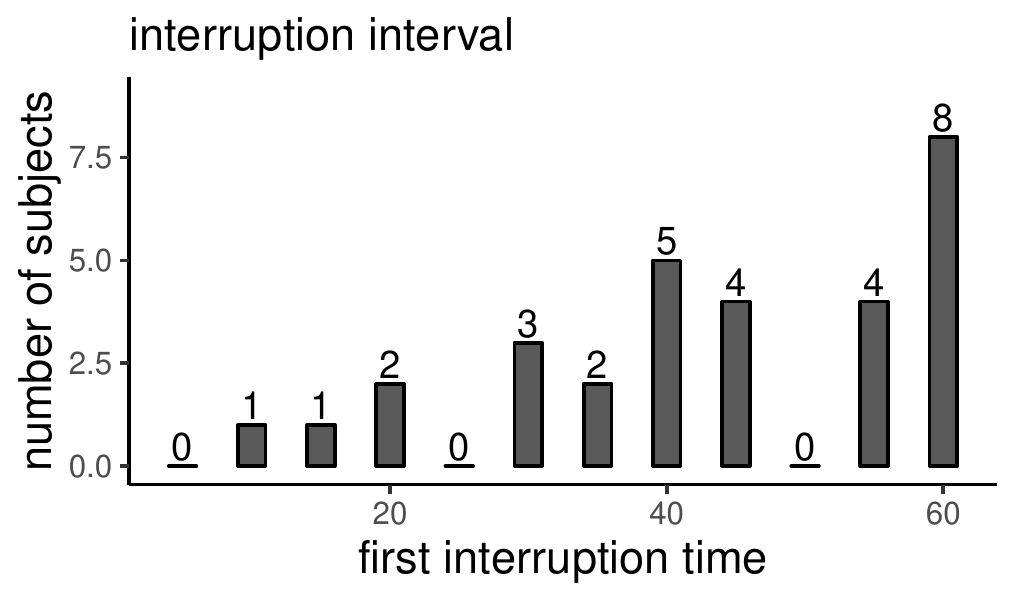}
\hspace{0.3cm}
\includegraphics[width=.3\linewidth]{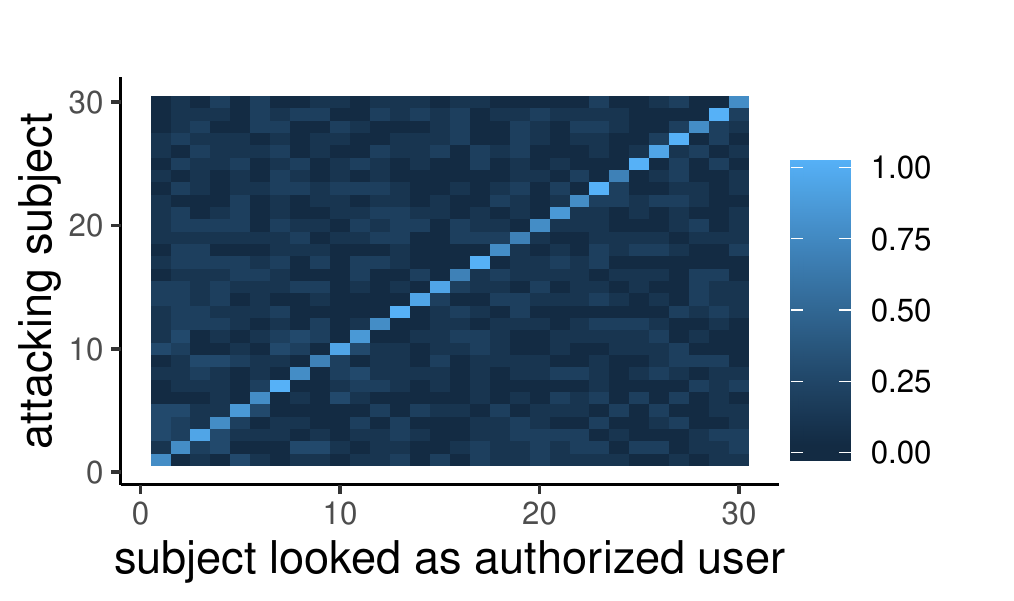}
\hspace{0.3cm}
\includegraphics[width=.3\linewidth]{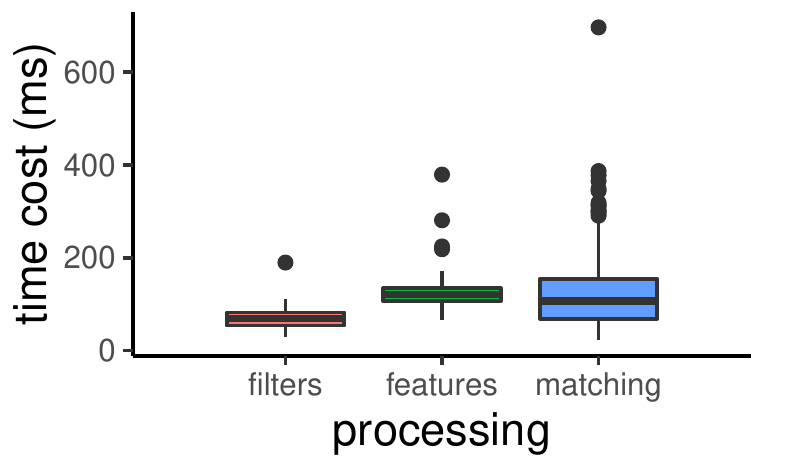}
\caption{Overall evaluation results of \name.~\textbf{(left)}: Time and frequency of first interruption happening;~\textbf{(middle)}: Confusion matrix on defending precision.~\textbf{(right)}: Time cost at 3 main processing stages, applying filters, extracting features and matching.}
\label{fig:primary}
\end{figure*}

\textbf{Mean interruption interval.} We first consider the case that \name authorizes legal users wrongly as adversaries~(true negative), which interrupts the user's operation due to the re-logging in. We ask 30 subjects to sit as in \Figure{deployment} for 60 minutes and record corresponding CSI based biometrics. We do continuous authentication every 5 minutes and record the time and frequency of the first interruption in \Figure{primary}~(left). For instance, \name first interrupts 5 subjects at time of the 40 minutes. Meanwhile, if \name does not interrupts in these 60 minutes, we record the first interruption time as the 60 minutes, that is, \name does not interrupts 8 subjects in these 60 minutes. We compute the mean interruption interval~($mI^2$) by using following equation.
\begin{equation}
   mI^2  = \sum_{t\in \left \{5,10,...,60\right\} }n_t \times t/N,
   \label{eq:mii}
\end{equation}
where $n_t$ is the \textit{amount} of the first interruption taking place at time $t$, and $N$ stands for the amount of subjects 30. Then, we have the average interruption interval of \name in the evaluation dataset is 43.5 minutes. We made a questionnaire about the acceptable interruption interval among these subjects, 27 out of 30 think this interruption interval is acceptable considering of the security issue.

\textbf{Mean authentication accuracy.} 
We examine the next metric, named mean authentication accuracy~($mA^2$), to evaluate \name performance on true positive~($TP$) authentication. As shown in \Figure{primary}~(left), one subject is interrupted by \name at the 10 minutes, which means \name works incorrectly at the second time on this subject (first time is at 5 minutes). Thus, we 
compute the accuracy of this situation as $(10-5)/10=50\%$. For example, if the first interruption happens at time 55, the corresponding accuracy is $(55-5)/55=90.91\%$. Note that, if \name does not interrupts a subject, the accuracy on this subject is 100\%. By this definition, we have the mean authentication accuracy as:
\begin{equation}
    mA^2 = \frac{ \sum_{t\in \left \{5,10,...,55\right\} }n_t \frac{t-5}{t}   + n_{60}\times 100\%}{N},
    \label{eq:maa}
\end{equation}
where the $n_{60}$ is the frequency of first interruption happening at the 60 minutes. Inputting the value shown in \Figure{primary}~(left), we have mean authentication accuracy as 88.16\%.

\textbf{Mean defending precision.}
We then evaluate the third metric, mean defending precision~($mDP$) of \name, which is a metric for  defending adversaries correctly. 

In particular, for one subject, we treat him/her as the authorized user, and consider the remaining 29 subjects as adversaries. We set \name does continuous authentication every 5 minutes, thus, every adversary is tested by $60/5=12$ times.
We repeat similar testings for the other 29 subjects.

As shown in \Figure{primary}~(middle), the value of the element in the $(i,j)$ block represents the frequency in our dataset that \name wrongly considers the $j^th$ adversary as authorized user when we doing above processing at the $i^th$ subject. Then we can compute mean defending precision with \Equation{mdp}.

\begin{equation}
    mDP = 1 - \frac{\sum_{i\neq j,~i,j\in[1,N]} p(i,j)}{(N-1)\times N}
    \label{eq:mdp}
\end{equation}
where $N$ is 30, $p(i,j)$ is the element value at the block of $(i,j)$. Finally, the mean defending precision of \name is 90.18\% based on \Figure{primary}~(middle).

\textbf{Authentication time delay.} The main authentication time delay consists of applying filters, computing CSI based biometrics and bimectrics matching. We use a desktop with Intel i5-3470S CPU and 32GB RAM to evaluate the authentication delay. We repeatedly do these computation and record the cost of time for 1K times, which results in a boxplot shown in \Figure{primary}~(right). From the figure, we know medians of time cost on these three processing stages are 70$ms$, 115$ms$ and 105$ms$, respectively. This light-weighted computation requirement enables \name run continuous authentication in real-time system.
We notice the maximal delay is 1300$ms$~(200+400+700), which is acceptable for the common usage.

\subsection{Micro-benchmark experiments}
There exist some empirical selections in designing \name algorithms. To make a better understanding on the relation between these selections and performance, we conduct micro-benchmark experiments in this subsection.

\textbf{Information reserving rate in dimensionality reduction.}
In the CSI based biometrics features processing, we apply PCA to reduce data dimensionality. 
In the overall evaluation, we selectively reserve 90\% information~(variance) of the data. Here we adjust the information reserving rate to evaluate the ability of dimensionality reduction.

\begin{table}[h]
	\begin{minipage}{1\linewidth}
		\centering
		\resizebox{7.5cm}{!}{
			\begin{tabular}{c||l|l|l}
				\hline
				\textbf{Reserved information} & $mI^2$ & $mA^2$ & $mDP$ \\ \hline\hline
				85\%        & 44.83   & 89.43\%   & 84.11\%   \\ \hline
				\textbf{90\%}        & \textbf{43.50}   & \textbf{88.16\%}   & \textbf{90.18\%}   \\ \hline
				95\%        &   41.50 & 84.49\%   & 92.23\%   \\ \hline
				100\%       & 37.80   & 81.46\%   & 83.50\%   \\ \hline
			\end{tabular}
		}
	\end{minipage}
	
	\begin{minipage}{1\linewidth}\vspace{0.3cm}
		\centering
		\resizebox{8cm}{!}{
			\begin{tabular}{c|c||l|l|l}
				\hline
				\textbf{Recording duration} & \textbf{Clustering group} & $mI^2$ & $mA^2$ & $mDP$ \\ \hline\hline
				1    & 4        & 36.67   & 80.16\%   & 92.34\%   \\ \hline
				\textbf{2}    & \textbf{4}        & \textbf{43.50}   & \textbf{88.16\%}   & \textbf{90.18\%}   \\ \hline
				3    & 4        & 45.17   & 89.83\%   & 90.46\%   \\ \hline\hline
				2    & 1        & 34.67   & 81.43\%   & 85.35\%   \\ \hline
				2    & 2        & 41.83   & 86.49\%   & 86.74\%   \\ \hline
				2    & 6        & 43.33   & 87.86\%   & 93.54\%   \\ \hline
				
			\end{tabular}
		}
	\end{minipage}
	\caption{Ablation results on empirical selection of, \textbf{up:}~reserved information rate by PCA; \textbf{bottom:}~CSI recording duration and clustering group for registering features. The result marked with bold font is the initial setting in \Sec{primary} }
	\label{tab:ablation}
\end{table}
As shown in \Table{ablation}~(up), we find (1) PCA is good for all three metrics; (2) if PCA is applied, reserving less information arises less interruption~(more authentication accuracy);  however, (3) if PCA is applied, reserving less information harms the precision of detecting adversaries. By analyzing this phenomena, we infer that 
major features of subjects are embedded in high variance dimensions, reserving these features helps to identify legal users continuously~(reason of 2).
Besides, we think the reason behind phenomena~(3) is a few subjects related features may exist in the small variance dimensions, and if we ignore them, \name works worse in identifying adversaries.

\textbf{CSI recording duration and clustering groups for preparing registering features.}
Having depicted in \Sec{design}, after authentic user logs in, \name records CSI series for a certain time to prepare registering CSI based biometrics features. In the primary evaluation, we record 2 minutes and divide them it 4 clustering groups, as illustrated in \Figure{classifier}~(left). We examine several other settings as shown in \Table{ablation}~(bottom).

From the first settings in \Table{ablation}~(bottom), we conclude increasing CSI recording duration can make user's features stable and lead to less interruptions and better performance on authorize legal users and defend illegal users. Meanwhile, we ascribe the reason of the last three settings results to the advantages of the multiple clustering strategy, depicted in \Sec{design} and illustrated in \Figure{classifier}.

\begin{figure}[t]
\hspace{-0.8cm}
\begin{minipage}{0.5\linewidth}
\centering
\includegraphics[width=.7\linewidth]{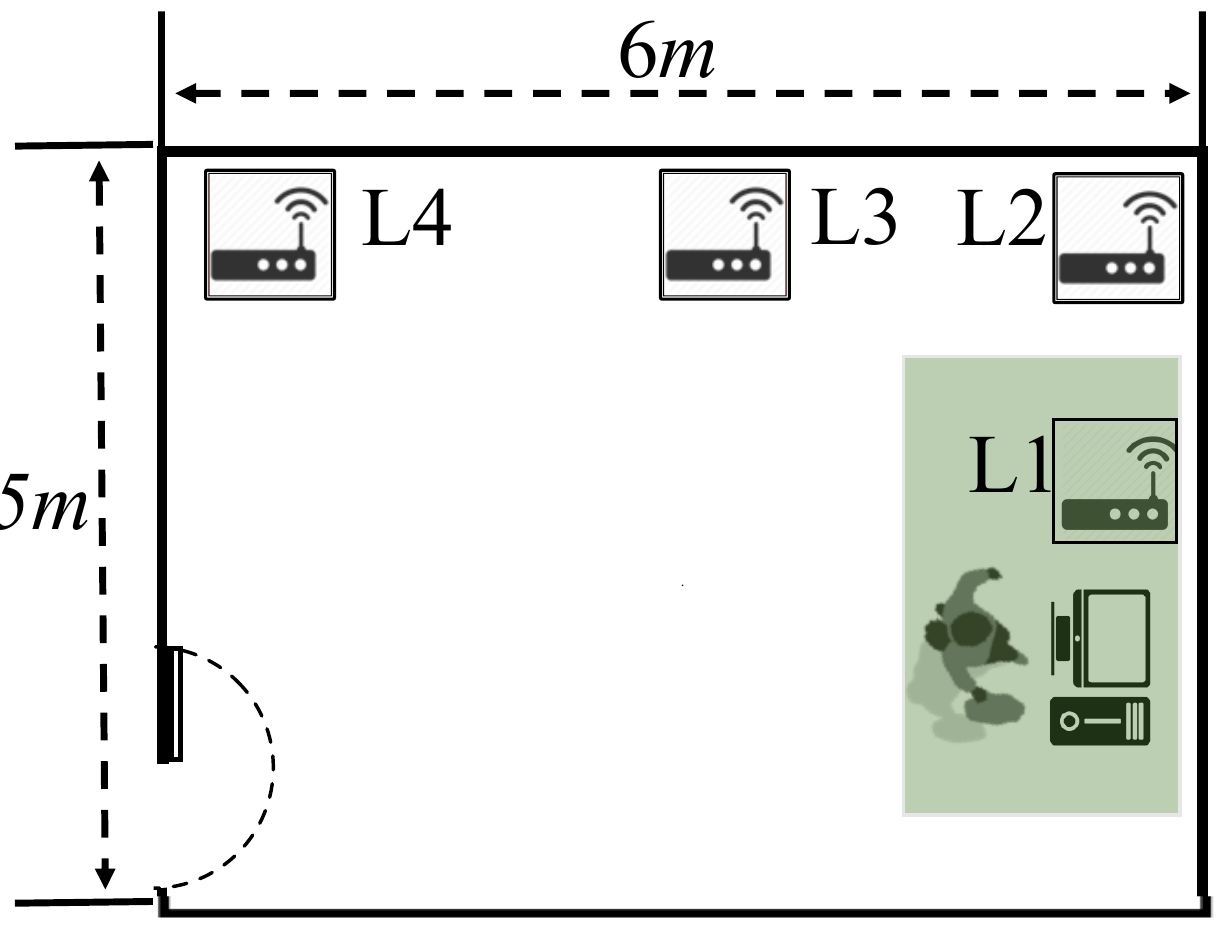}
\end{minipage}
\hspace{-0.6cm}
\begin{minipage}{0.5\linewidth}
\centering
\resizebox{5.5cm}{!}{
\begin{tabular}{c||l|l|l|l}
\hline
 \textbf{Location} & L1 & L2 & L3 & L4 \\ \hline\hline
Tx-Rx distance & 0.8$m$   & 2.8$m$   & 3.5$m$  & 5.5$m$ \\\hline
User position      & NLOS   & NLOS   & LOS  &LOS  \\\hline
$mI^2$        & 43.50   & 40.67   & 41.33 & 39.17    \\\hline
$mA^2$       & 88.16\%   & 85.74\%   & 86.85\%  & 84.84\% \\\hline
$mDP$     & 90.18\%   & 88.18\%   & 89.27\%& 87.24\% \\\hline

\end{tabular}
}
\end{minipage}
\caption{Settings on relative location among transmitter, receiver and user. The setting marked with green shadow is the data collecting scene shown in \Figure{deployment}.}
\label{fig:location}
\end{figure}

\textbf{Relative location among transmitter, receiver and user.}
We change router position at 4 typical places in a $5m\times6m$ room to evaluate \name, which leads to users sitting in line-of-sight~(LOS) and non-line-of-sight~(NLOS), shown in \Figure{location}~(left). The scene of the overall evaluation is marked with green shadow. The involved subjects and process of data collection keep the same with the overall evaluation. 

The results shown in \Figure{location}~(right) demonstrate \name is robust to the change of relative location among transmitter, receiver and user, which is practical for use. Specially, comparing results of L1/L2, L3/L4, we notice \name works better if transmitter and receiver put closer. Meanwhile, comparing results of L2 and L3, which with similar distance, we conclude that \name works better when users sitting in LOS. 

\textbf{Interference from other subjects.}
All above results are derived from situation that only the user is in the room shown in \Figure{location}~(left), which arise our concerns on applying \name in a more normal situation. Next, we evaluate the performance of \name when facing the interference from other subjects in surroundings. The data collection process and evaluation metrics in this part are much different to those in the above, thus, we explain them in details before going to results. Note that, in this part, the relative location is the same as the L1 shown in \Figure{location}~(left) if not mentioned. 

\begin{figure}[t]
\centering
\includegraphics[width=0.98\linewidth]{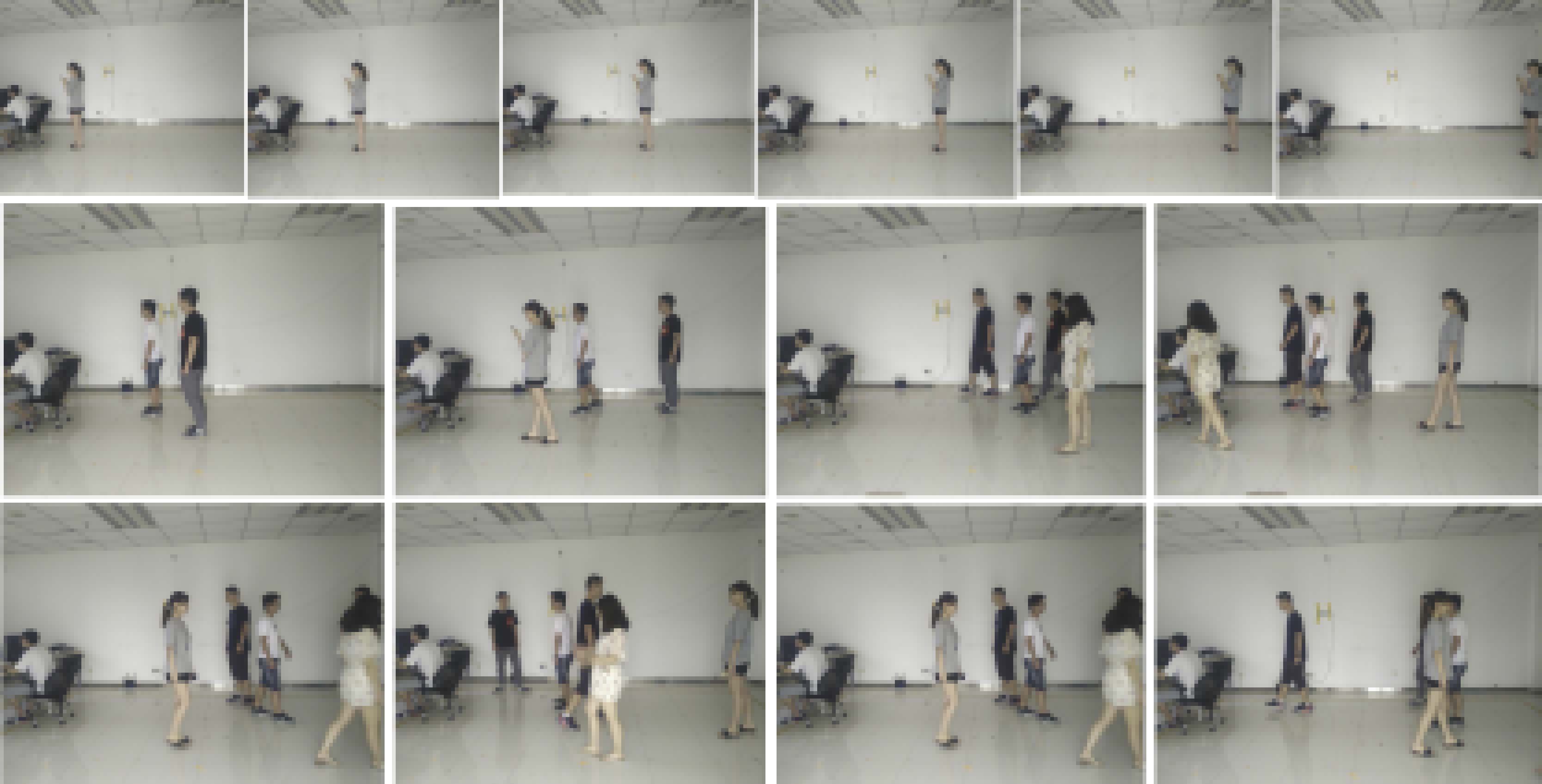}
\caption{Testing influence from nearby subjects.
\textbf{(top)}: various distances to the user;
\textbf{(middle)}: increasing nearby subjects' amount;
\textbf{(bottom)}: influence from subjects' casual motions.
}
\label{fig:other}
\end{figure}

\textit{1) Distance of one other subject}. Illustrating in the first row of \Figure{other}, we first asked one user to sit before a monitor and collected corresponding CSI series for 2 minutes, then we asked one subject to stand behind the use with a distance about 0.6$m$ and collected corresponding CSI series for 2 minutes. During CSI collection, the user was asked to do the least motions. The former 2-minute series are to train classifier, and the later 2-minute series are for testing \name when facing other subject. We tested distances around 0.6$m$, 1.2$m$, 1.8$m$, 2.4$m$, 3.0$m$ and 3.6$m$ and did this on up to 10 users. 

To make it clear, we divide the later 2-minute series into 10 testing samples, then we obtain 100 testing samples on 10 users for every distance. The authentication accuracy are 73\%, 81\%, 87\%, 90\%, 93\%, 91\%, respectively. This indicates \name still works well when facing the interference from a subject 1.8$m$ away from this relative location.

\textit{2) Number of other subjects}. As shown in the \Figure{other}~(middle), we first asked user to sit before a monitor and collected corresponding CSI series for 2 minutes, then, we asked 2 other subjects to stand behind the user and collected CSI series for 2 minutes. Subjects were asked to change their positions randomly for 10 times, and number of tested subjects increase from 2 to 5. Thus, we have $10\times10=100$ samples when testing every specific number. The authentication results are 87\%, 83\%, 80\%, and 75\% for number of 2, 3, 4 and 5, respectively. This indicates user may have to re-log in with his/her keys such as password, fingerprint, face \etc if many subjects appear in surrounding suddenly. 

\textit{3) Motions of other subjects.} We first asked 5 subjects to move casually in the room, then one user was asked to sit before the monitor. Concurrently, we recorded CSI series for 30 minutes for training classifiers. The amount of involved users is still 10. We use the first 2-minute CSI series to train classifiers, meanwhile, \name is set to do continuous authentication every 3 minutes. Similar to metrics in overall evaluation, finally, we have $mI^2$, $mA^2$ and $mDP$ of 17.70, 82.07\% and 84.23\%, which indicates \name can still work properly in noisy environment.  

\textit{ 4) Relative location of other subjects}. 
Please look at the \Figure{location}~(left), in the above three experiments, we selected relative position setting of L1 and asked other subjects appearing behind the user, which cause the interference of other subjects is mainly from NLOS. To test the interference from LOS, we applied setting of L4 and utilized metrics as the above $3^{rd}$ experiment. Not surprisingly, the performance decreases to $mI^2$ of 12.60, $mA^2$ of 67.07\% and $mDP$ 69.50\%, respectively. This problem matches the human body impacts on Wi-Fi signals depicted in \Sec{model}. We highly recommend to set relative position of transmitter and receiver there where would cause the least LOS interference from other subjects.

\subsection{Comparison evaluation}

We make an extended evaluation to test the possibility of applying \name as an alternative log-in authentication keys like fingerprint, face, \etc. 
To do this, we use the dataset collected at the \Sec{primary} to train multi-class classifiers with LibSVM\cite{CC01a}~(radius basis function kernel, L1 regularization, L1 loss and one-against-all strategy). For each subject, data collected in the first 48 minutes is for training, the remaining 12-minute data is for testing. 

\begin{figure}[th]
\centering
\includegraphics[width=.48\linewidth]{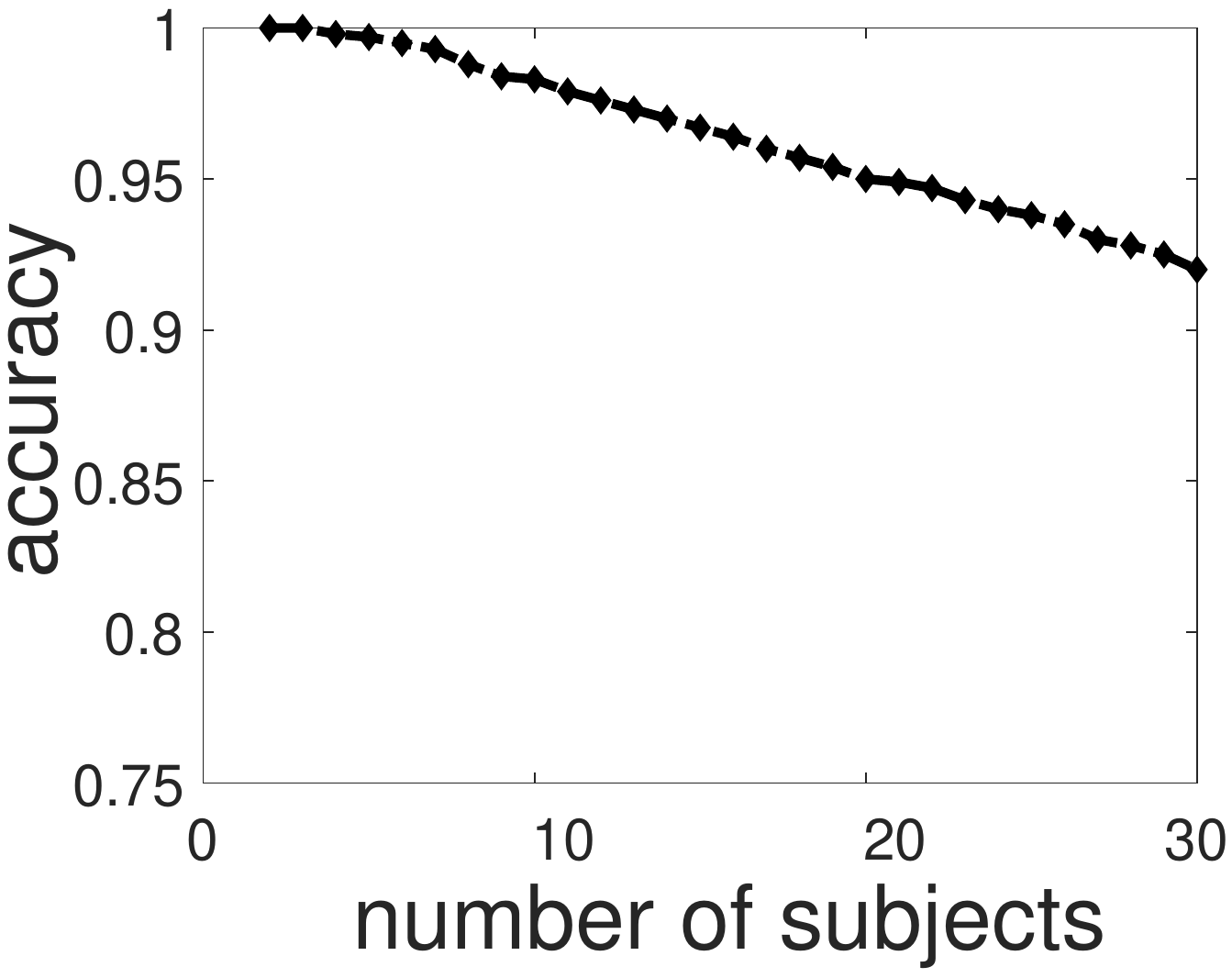}
\includegraphics[width=.48\linewidth]{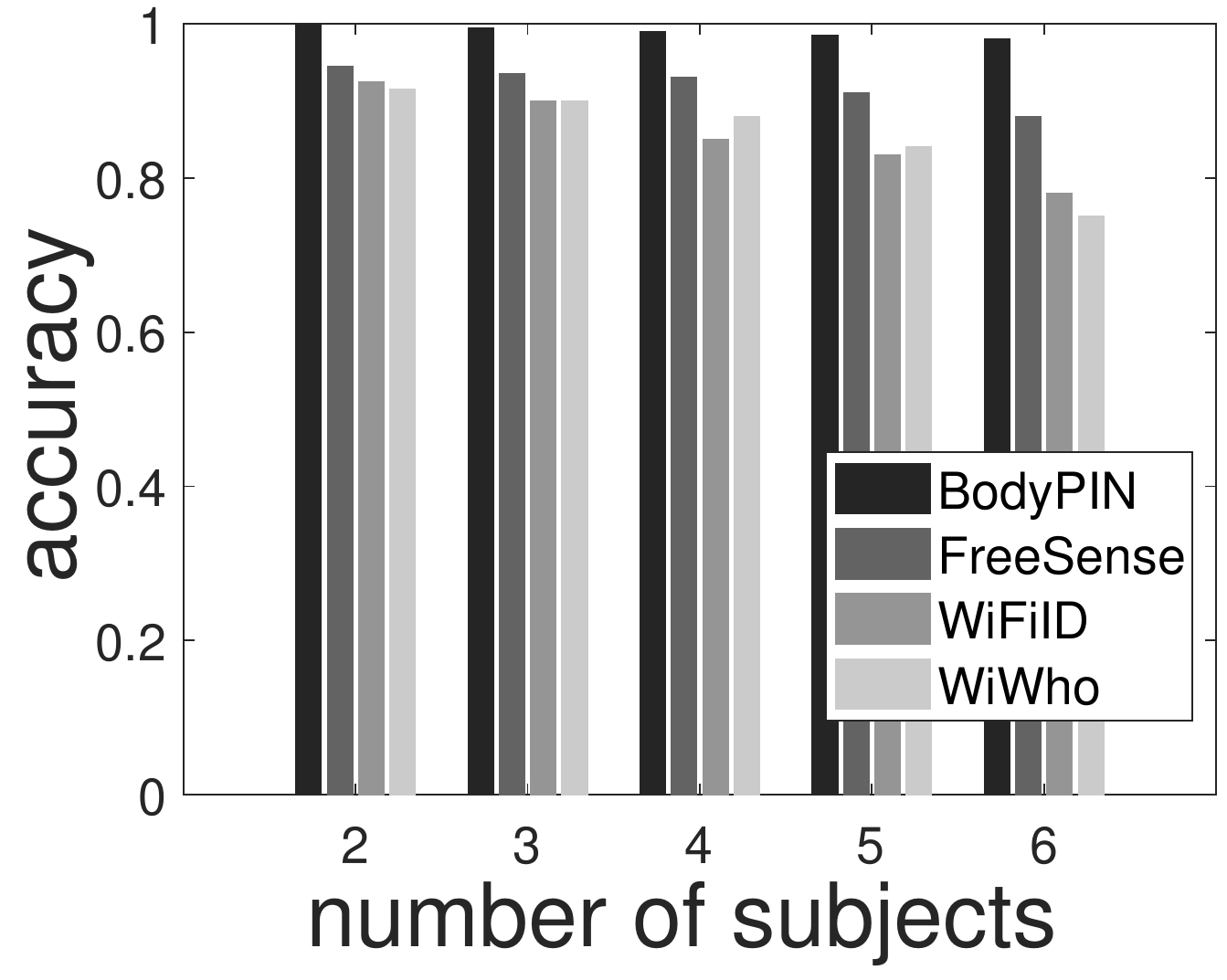}
\caption{\name is extended as multi-class classifiers on CSI based biometrics of 30 subjects. \textbf{(left)}: classification accuracy; \textbf{(right)}: comparison with existing works.}
\label{fig:addacc}
\end{figure}

We evaluate the user capacity of \name from 2 to 30. When evaluating the user capacity of $i~(i\in[2,30])$, we randomly select training dataset of $i$ subjects and test with their testing sets. 
As shown in \Figure{addacc}~(left), \name can achieve good performance, \eg nearly 100\% accuracy with few subjects and more than 92\% accuracy with the user capacity of 30. We compare the accuracy with FreeSense\cite{xin2016freesense}, WiFiID\cite{zhang2016wifi} and WiWho\cite{zeng2016wiwho}, and find \name outperforms them, shown in \Figure{addacc}~(right). 

The additional evaluation demonstrates a possibility that applying \name as an alternative log-in authentication keys. However, it is still an open problem to make CSI based biometrics stable and accurate as high as fingerprint, face, \etc.

\section{Related Work}\label{sec:related}

\textbf{Bio-eletromagnetics}. The \systemname design relates to the bio-eletromagnetics literatures \cite{christ2006characterization,dove2014analysis,melia2013electromagnetic,gabriel1996dielectric2}. Some human tissues, such as body muscle, kidney and liver, with different dielectric properties 
are measured by signals from 10Hz to 20GHz \cite{gabriel1996dielectric2}. The body's absorption is studied by \cite{christ2006characterization} in range of 30MHz to 6GHz, and \cite{melia2013electromagnetic} in range of 1GHz to 15GHz. In \cite{dove2014analysis}, an in-body electromagnetic transmit model is proposed and tested in 2.45GHz. These works validate our body could have unique impacts on wireless signals. Based on this, we further use the effective body model from this domain in the \systemname design.

\textbf{Biometrics based continuous authentication}. The camera can achieve continuous authentication, \eg sensing the user's cloth and skin~\cite{niinuma2010continuous} 
, and the gaze moving patterns \cite{eberz2015preventing}. As stated in the introduction, it requires strict line of the sight and lighting conditions. More importantly, it may have severe privacy concerns about the recorded video. To overcome these issues, there are recent designs using wireless to achieve the continuous authentication, like \cite{lin2017cardiac}, which however requires a dedicated hardware design. Compared with these existing works, \systemname is a wireless-based solution avoiding camera's drawbacks, while utilizes commercial Wi-Fi devices only.

\textbf{Wi-Fi based human identification}. There also exist many Wi-Fi-based human identification systems, \eg WiWho\cite{zeng2016wiwho}, WiFi-ID\cite{zhang2016wifi}, FreeSense\cite{xin2016freesense}
and Radio-Bio\cite{xu2017radio}. However, they require the user to perform certain activities, \eg walking, as they essentially recognize the user's activities, instead of the users themselves. Therefore, these designs are not suitable for the continuous authentication, since frequently performing the required activities could easily interrupt the user's normal usage of the computer system and dramatically sacrifices the user experience.

\textbf{Wi-Fi time series matching}. Techniques of Wi-Fi time series matching are also related to this paper. Existing techniques mainly fall into two categories. First, the dynamic time wrapping\cite{berndt1994using} is widely used in Wi-Fi time series comparison for action recognition~\cite{ali2015keystroke,palipana2018falldefi}
Another category is to convert the Wi-Fi time series to statistics features such as \textit{minimum}, \textit{maximum} and \textit{mean} in 
\cite{zeng2016wiwho,zhang2016wifi,xin2016freesense,xu2017radio}. 
Guided by the second category, we also extract useful features in \systemname, while our feature extraction is inspired by the bio-eletromagnetic model derived, so that they can uniquely and reliably represent different users for the continuous authentication.

\section{Conclusion}\label{sec:conclusion}
In this paper, we demonstrate a contactless continuous authentication system, \name, by using the human body biometrics features conveyed in Wi-Fi signals. \name requires no extra or dedicated  wireless hardware but achieves promising authentication performances, \ie acceptable interruption interval, high authenticating  and defending accuracy, lightweight computation, resilience on surrounding people, \etc. Due to these strengths, we believe \name could be a useful and practical system.

\bibliographystyle{IEEEtran}
\bibliography{bibo1}

\end{document}